\documentclass[pdflatex,sn-mathphys,referee]{sn-jnl}
\usepackage{siunitx}
\usepackage{physics}
\usepackage[capitalise]{cleveref}

\jyear{2023}
\unnumbered

\newcommand{\VLM}{V_\mathrm{LP}}
\newcommand{\VRM}{V_\mathrm{RP}}
\newcommand{\VLI}{V_\mathrm{LT}}
\newcommand{\VRI}{V_\mathrm{RT}}
\newcommand{\VH}{V_\mathrm{H}}
\newcommand{\VL}{V_\mathrm{L}}
\newcommand{\VR}{V_\mathrm{R}}
\newcommand{\IL}{I_\mathrm{L}}
\newcommand{\IR}{I_\mathrm{R}}
\newcommand{\GLL}{G_\mathrm{LL}}
\newcommand{\GRR}{G_\mathrm{RR}}
\newcommand{\GLR}{G_\mathrm{LR}}
\newcommand{\GRL}{G_\mathrm{RL}}

\newcommand{\su}{\uparrow} 
\newcommand{\sd}{\downarrow} 
\newcommand{\bpm}{\begin{pmatrix}}
\newcommand{\epm}{\end{pmatrix}}

\newcommand{\dg}{^{\dagger}}

\newcommand{\nn}{\nonumber \\} 

\newcommand{\muL}{\mu_\mathrm{L}}
\newcommand{\muR}{\mu_\mathrm{R}}
\newcommand{\muH}{\mu_\mathrm{H}}

\newcommand{\Ez}{E_\mathrm{Z}}
\newcommand{\Ezh}{E_\mathrm{ZH}}

\newcommand{\Egap}{E_\mathrm{gap}}

\newcommand{\tso}{w_\mathrm{SO}}
\begin{document}

\title{Robust poor man's Majorana zero modes using Yu-Shiba-Rusinov states}

\author[1]{\fnm{Francesco}~\sur{Zatelli}}
\equalcont{These authors contributed equally to this work.}

\author[1]{\fnm{David}~\sur{van Driel}}
\equalcont{These authors contributed equally to this work.}

\author[1]{\fnm{Di}~\sur{Xu}}
\equalcont{These authors contributed equally to this work.}

\author[1]{\fnm{Guanzhong}~\sur{Wang}}
\equalcont{These authors contributed equally to this work.}

\author[1]{\fnm{Chun-Xiao}~\sur{Liu}}

\author[1]{\fnm{Alberto}~\sur{Bordin}}

\author[1]{\fnm{Bart}~\sur{Roovers}}

\author[1]{\fnm{Grzegorz~P.}~\sur{Mazur}}

\author[1]{\fnm{Nick}~\sur{van~Loo}}

\author[1]{\fnm{Jan~C.}~\sur{Wolff}}

\author[1]{\fnm{A.~Mert}~\sur{Bozkurt}}

\author[2]{\fnm{Ghada}~\sur{Badawy}}

\author[2]{\fnm{Sasa}~\sur{Gazibegovic}}

\author[2]{\fnm{Erik~P.~A.~M.}~\sur{Bakkers}}

\author[1]{\fnm{Michael}~\sur{Wimmer}}

\author*[1]{\fnm{Leo~P.}~\sur{Kouwenhoven}}\email{l.p.kouwenhoven@tudelft.nl}

\author[1]{\fnm{Tom}~\sur{Dvir}}

\affil[1]{\orgdiv{QuTech and Kavli Institute of NanoScience}, \orgname{Delft University of Technology}, \postcode{2600 GA} \orgaddress{\city{Delft}, \country{The Netherlands}}}

\affil[2]{\orgdiv{Department of Applied Physics}, \orgname{Eindhoven University of Technology}, \postcode{5600 MB} \orgaddress{\city{Eindhoven}, \country{The Netherlands}}}

\date{\today}

\abstract{
The recent realization of a two-site Kitaev chain featuring ``poor man's Majorana'' states demonstrates a path forward in the field of topological superconductivity.
Harnessing the potential of these states for quantum information processing, however, requires increasing their robustness to external perturbations.
Here, we form a two-site Kitaev chain using proximitized quantum dots hosting Yu-Shiba-Rusinov states.
The strong hybridization between such states and the superconductor enables the creation of poor man's Majorana states with a gap larger than $\SI{70}{\micro eV}$.
It also greatly reduces the charge dispersion compared to Kitaev chains made with non-proximitized quantum dots.
The large gap and reduced sensitivity to charge fluctuations will benefit qubit manipulation and demonstration of non-abelian physics using poor man's Majorana states.
}

\maketitle

\clearpage

Kitaev chains based on quantum dots (QDs) coupled via a hybrid semiconductor-superconductor heterostructure are a promising avenue for the creation of Majorana bound states \cite{Kitaev.2001,Sau.2012}.
Even a minimal chain, consisting of only two QDs, supports fine-tuned Majorana zero modes known as ``poor man's Majoranas'' (PMMs)~\cite{Leijnse.2012}.
These PMM states do not benefit from topological protection, but already exhibit robustness to local perturbations and quadratic protection from global fluctuations in the chemical potential~\cite{Leijnse.2012}.
Moreover, PMM states obey non-abelian exchange statistics, thus providing a favourable platform for braiding and Majorana-based qubit experiments in the near future \cite{Liu.2022.Fusion,Boross.2023,Tsintzis.2023,Pino.2023}. 
In Ref.~\cite{Dvir.2023}, we have recently realized the two-site Kitaev chain by coupling two QDs formed in an InSb/Al hybrid nanowire.
While providing the necessary proof of concept, the gap separating PMMs from excited states is not much higher than the electron temperature. 
Furthermore, appreciable noise in the measured spectra indicate sensitivity to charge fluctuations, making practical use of such states infeasible.

Insensitivity to charge noise has been successfully achieved in transmon qubits by increasing the ratio between superconducting coupling (Josephson energy) and charging energy, thereby suppressing the charge dispersion of the spectrum~\cite{Koch.2007,schreier2008Suppressing}.
We follow a similar approach by strongly coupling our QDs to the superconductor.
This induces superconducting correlations on the QDs leading to energy eigenstates known as Yu-Shiba-Rusinov (YSR) states \cite{Yu1965bound,shiba1968classical,rusinov1969superconductivity,Buitelaar.2002,Bauer.2007, meng2009self, GroveRasmussen.2009,Chang.2013,Jellinggaard.2016,Grove-Rasmussen.2018}.
Spin-polarized YSR states can also serve as the sites of a Kitaev chain, as investigated theoretically~\cite{Fulga.2013,Nadj-Perge.2013,Pientka.2015,schecter2016self,Liu.2022,Tsintzis.2022} and experimentally using scanning tunneling microscopy~\cite{NadjPerge2014,ruby2015end,Jeon.2017,schneider2021Topological,schneider2022Precursors}.
In this work, we realize a two-site Kitaev chain using YSR states in a hybrid semiconducting InSb nanowire partly covered with a superconducting Al film.
We show that the sensitivity of the resulting PMMs to charge fluctuations affecting the QDs decreases by two orders of magnitude compared to a QD-based Kitaev chain~\cite{Dvir.2023}.
In addition, we measure a gap between ground and excited states of $\Egap =\SI{76}{\micro eV}$, a threefold increase compared to our previous report~\cite{Dvir.2023}.
Complementary results are also reported in a parallel study on a 2D hybrid platform using an InAsSb/Al two-dimensional electron gas~\cite{tenHaaf.2023}, demonstrating the wider applicability of our approach. 

\begin{figure*}[ht!]
    \centering
    \includegraphics[width=\textwidth]{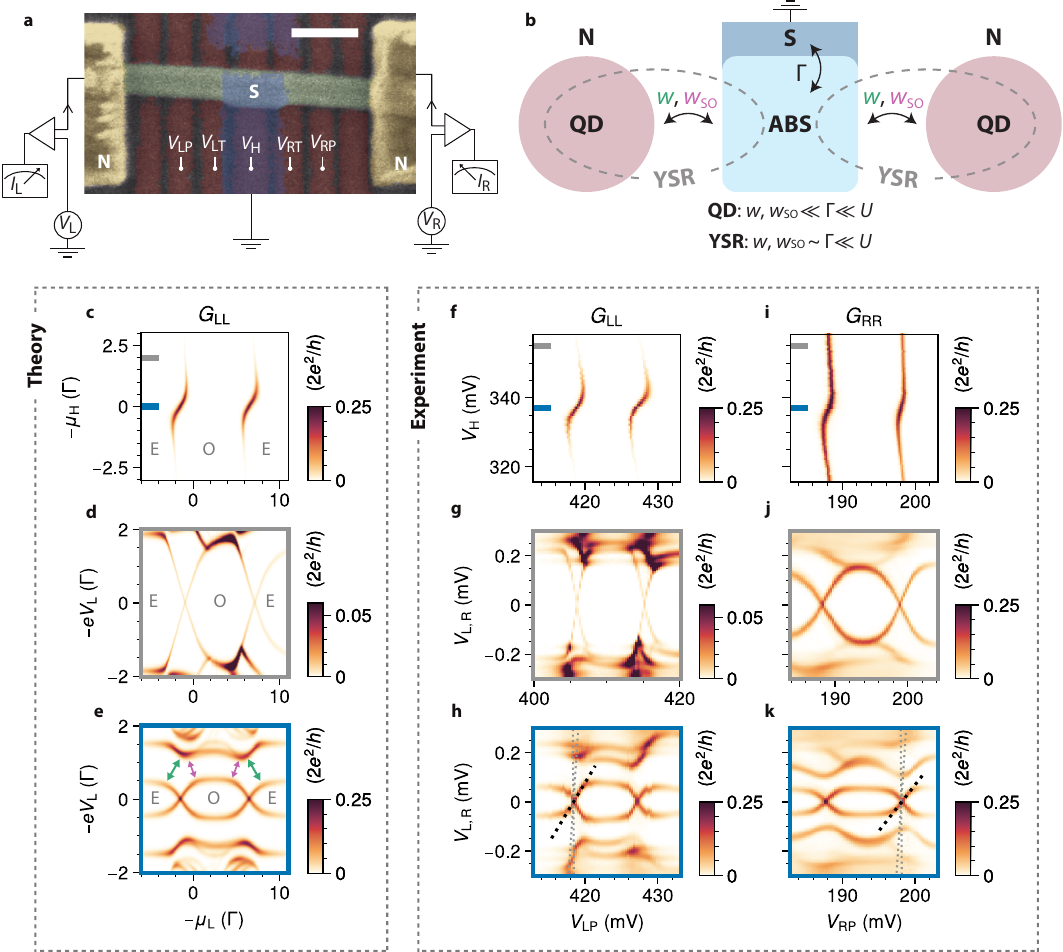}
    \caption{\textbf{YSR states formed by hybridizing QDs with an ABS.}
    \textbf{a,} False-colored scanning electron micrograph and measurement circuit of the device. Scale bar is $\SI{200}{nm}$.
    \textbf{b,} Illustration of the QD-ABS-QD model. Two QDs are coupled to the same ABS via spin-conserving and spin-flipping tunneling. QD and YSR regimes are defined by the relative energy scales shown below the sketch.
    \textbf{c,} Numerical CSD using zero-bias conductance of a QD coupled to an ABS. ``E''/``O'' indicate even/odd occupation of the QD.
    \textbf{d, e,} Numerical conductance spectrum of a QD coupled to an ABS as a function of QD chemical potential when ABS is far (d) and close to its energy minimum (e). In e, green/pink arrows indicate avoided crossings due to spin-conserving/spin-flipping tunneling.
    \textbf{f--h,} Same as c--e for the left QD coupled to an ABS, experimentally measured using local conductance. Gray/blue ticks in f mark $\VH$ values at which g/h are measured. The right QD is off resonance. In h, we extract a lever arm $\alpha \approx 0.05e$ by fitting the spectrum with the black dotted line at the zero-energy crossing. This procedure overestimates the actual lever arm since it does not account for capacitance to the normal lead.  For comparison, we plot the profile of a typical Coulomb diamond in our device with gray dotted lines (see \cref{ED2:QDs_highbias}a).
    The $\VLM$ range in g is shifted because of a gate jump affecting the left QD.
    \textbf{i--k,} Same as f--h, but for the right QD. The extracted lever arm is $\alpha \approx0.04e$.
}
    \label{fig:1}
\end{figure*}

\section{Fabrication and setup}

\cref{fig:1}a shows a scanning electron microscope image of the reported device.
It consists of an InSb nanowire (green) placed on top of a series of bottom gates \cite{Badawy.2019}.
The middle part of the nanowire is covered by a thin Al shell (blue), forming a superconductor-semiconductor hybrid whose electrochemical potential is controlled by a plunger gate ($\VH$).
On both sides of the hybrid segment, QDs are formed in the nanowire using three gates each.
The electrochemical potential of each QD is controlled by a plunger gate ($\VLM$ and $\VRM$ for the left and right QDs, respectively), and the couplings between the QDs and the hybrid are controlled by tunnel gates ($\VLI$ and $\VRI$ for the left and right QDs, respectively).
A normal lead is attached to each QD, separated by another gate-defined tunnel barrier. The superconducting lead is kept grounded at all times. In addition, the two normal leads are connected to off-chip multiplexed resonators for fast RF reflectometry measurements~\cite{Hornibrook.2014}, using the setup described in Ref.~\cite{wang2022parametric}.
Each lead is voltage biased independently with respect to the grounded Al, with voltages $\VL$ and $\VR$ on the left and right leads, respectively.
The currents ($\IL$ and $\IR$ on the left and right leads), the local conductances ($\GLL= d\IL/d\VL$, $\GRR= d\IR/d\VR$), and the non-local conductances ($\GRL= d\IR/d\VL$, $\GLR= d\IL/d\VR$) are measured simultaneously. When the full conductance matrix is measured, we correct for line resistance as described in the Methods section. Further fabrication and setup details can be found in our previous publications~\cite{Heedt.2021, Borsoi.2021, Mazur.2022}.
The experiment is conducted in a dilution refrigerator with a base temperature of $\SI{30}{mK}$. 
A magnetic field of $B = \SI{150}{mT}$ is applied along the nanowire axis, inducing a Zeeman splitting of approximately \SI{200}{\micro eV} in the QDs (\cref{ED3:QDs_sweetspot}).

\section{YSR states in quantum dots}

We model our system using a three-site model in which the hybrid is considered as a single Andreev bound state (ABS) in the atomic limit~\cite{Bauer.2007,meng2009self} tunnel-coupled to two QDs (see schematics in \cref{fig:1}b) \cite{Tsintzis.2022}.
The QDs have charging energy $U$, Zeeman splitting $\Ez$, and chemical potentials $\muL, \muR$ for the left and right QD, respectively.
The ABS has an induced gap $\Gamma$, which in the atomic limit \cite{Bauer.2007,meng2009self} can be identified with its tunnel coupling to the bulk superconductor.
Its charging energy is negligible due to the screening of the grounded Al film.
It also has Zeeman splitting $\Ezh$, which is smaller than that of the QDs due to metallization of the ABS~\cite{Reeg.2018}. 
We ensure $\Ezh<\Gamma$ so that the ground state of the ABS is always a BCS singlet.
The electron-hole composition of the ABS depends on its chemical potential $\muH$.

In our model, the QDs are coupled to the ABS by spin-conserving and spin-flipping tunneling due to spin-orbit interaction, with amplitudes $w$ and $\tso$, respectively~\cite{Tsintzis.2022}.
The hybridization between QDs and the ABS becomes significant when $w,\tso \sim \Gamma$. 
As a consequence, the QDs become proximitized and form YSR states~\cite{Grove-Rasmussen.2018,Liu.2023}, which we distinguish from ABSs because of their large charging energy \cite{Bauer.2007, meng2009self, GroveRasmussen.2009, Chang.2013, kirsanskas2015YuShibaRusinov, Jellinggaard.2016}.
We refer readers to the Supplementary Information and the parallel work of Ref.~\cite{Liu.2023} for theory models of the strong coupling regime investigated in this work.

To understand the nature of the YSR states formed in proximitized QDs, we first examine the coupling between a single QD and an ABS in the hybrid segment, following Ref.~\cite{Grove-Rasmussen.2018}.
\cref{fig:1}c shows the theoretical zero-bias conductance of a QD-ABS charge stability diagram (CSD), while the second QD is off-resonance.
The two vertical features indicate parity transitions of the system, largely corresponding to the consecutive filling of a single orbital of the QD by two electrons.
The S-shaped conductance features result from renormalization of the QD energy via hybridization with the ABS.
The ABS reaches its energy minimum at charge neutrality, i.e., $\muH=0$, where its excitation is equal-parts electron and hole.
The hybridization with the QD is also maximal here due to their minimal energy separation, evident in the enhanced zero-bias conductance as local Andreev reflection becomes stronger \cite{Grove-Rasmussen.2018,Liu.2023}.
The QD spectra in panels~d and e further reveal its hybridization with the ABS. 
When the latter is away from charge neutrality, the QD spectrum as a function of $\muL$ (\cref{fig:1}d) exhibits straight features reminiscent of Coulomb diamonds. 
As the ABS approaches $\muH=0$ (\cref{fig:1}e), such features evolve into an eye-shaped spectrum typical of YSR states \cite{Pillet.2010,Lee.2014,Jellinggaard.2016,Scherbl2020,Scherbl2022}.
The arrows indicate avoided crossings in the excited states of the spectrum produced by spin-conserving (green) and spin-flipping (pink) tunneling between the ABS and the QD levels~\cite{Liu.2023}.

The hybrid segment of our device features multiple discrete ABSs well-separated from each other in $\VH$ (\cref{ED1:hybrid_spectrum}).
We operate in a $\VH$ range containing a single ABS level. 
In \cref{fig:1}f, we show the zero-bias conductance measured with the left lead as a function of $\VLM$ and $\VH$. $\VRM$ is fixed to keep the right QD off-resonance.
We observe the QD-ABS charge stability diagram features described in panel c, indicating the presence of an ABS reaching its energy minimum at $\VH\approx\SI{337}{mV}$.
The QD-ABS hybridization is further confirmed by the QD spectrum being in agreement with the model when the ABS is away from (\cref{fig:1}g) and at its energy minimum (\cref{fig:1}h).

The YSR zero-energy excitations in \cref{fig:1}h are our building blocks of a Kitaev chain.
Compared to a non-proximitized QD zero-energy crossing, these YSR crossings have noticeably weaker energy dispersion as a function of gate (see dashed lines in panel~h).
To quantify this observation, we can estimate the lever arm of the YSR excitation at charge degeneracy using its gate-dispersion slope: $\alpha \equiv \partial E / \partial \VLM$.
In contrast to the above-gap QD lever arm of $\alpha \approx 0.4e$ in our devices (\cref{ED2:QDs_highbias}a), this subgap lever arm reduces to $\alpha \approx 0.2e$ (\cref{ED2:QDs_highbias}b) when the ABS is detuned from its charge neutrality. 
Tuning the ABS to its energy minimum further reduces the lever arm to $\alpha \approx 0.05e$ (\cref{fig:1}h).
This signals a strong reduction in the effective charge of the fermionic excitation, attributable to charging energy renormalization, QD-ABS hybridization, and electron-hole superposition \cite{Schindele.2014,gramich2017Andreev}, as detailed in \cref{ED:th_QDABS_wf}.

The right QD shows similar behavior when coupled to the same ABS (\cref{fig:1}i--k). 
However, when the ABS is away from charge neutrality, the subgap conductance is not significantly suppressed (\cref{fig:1}i) and the QD spectrum still shows typical YSR features (\cref{fig:1}h), albeit with a small superconducting coupling. 
This could be due to residual proximity resulting from direct coupling between the QD and the superconducting film since $\VH$ does not affect it appreciably (\cref{ED11:QDABS_right}).

\section{Coupled YSR states}

\begin{figure*}
    \centering
    \includegraphics[width=0.33\textwidth]{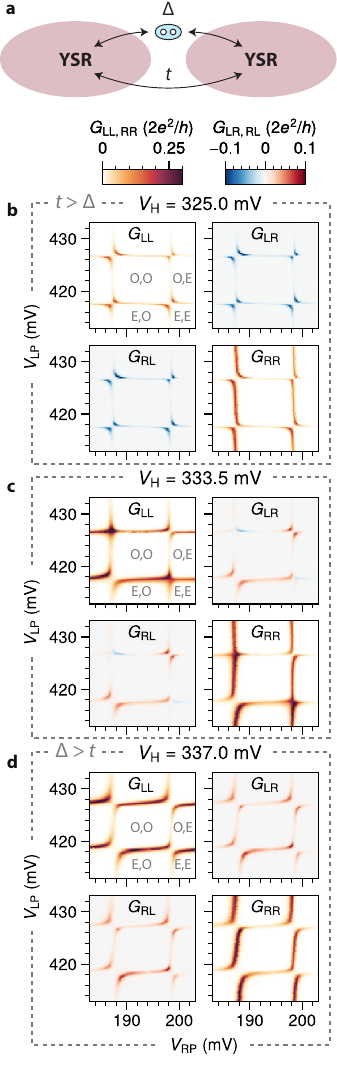}
    \caption{\textbf{ECT and CAR coupling between YSR states.} 
    \textbf{a,} Illustration of the effective two-site system of YSR states coupled via ECT ($t$) and CAR ($\Delta$).
    \textbf{b--d,} Conductance matrices of CSDs measured at different $\VH$. By tuning the electrochemical potential of the ABS in the hybrid, it is possible to continuously vary $t$ and $\Delta$. In b, avoided crossings along the anti-diagonal are observed, indicating $t>\Delta$.
    The opposite regime is shown in d, with all the avoided crossings along the diagonal, indicating $t < \Delta$. The crossover between these two regimes is shown in c, where two crossings indicate $t \approx \Delta$.
    }
    \label{fig:2}
\end{figure*}

Recent theoretical \cite{Liu.2022,Tsintzis.2022} and experimental \cite{Wang.2022,Dvir.2023,Bordin.2022} works have shown that an ABS can mediate elastic co-tunneling (ECT) and crossed Andreev reflection (CAR) between QDs. 
These two processes implement the hopping and pairing terms of the original Kitaev chain model \cite{Kitaev.2001,Sau.2012}. 
To form PMM states in a two-site chain, the amplitudes of both terms must be equal~\cite{Leijnse.2012}. 
Such control can be achieved by tuning the electrochemical potential of the ABS in the hybrid nanowire \cite{Liu.2022,Tsintzis.2022,Dvir.2023,Bordin.2022}.

Similar effective ECT and CAR couplings, with respective amplitudes $t$ and $\Delta$, also emerge between YSR states (\cref{fig:2}a) formed by strongly coupling QDs to the same ABS \cite{Trocha2015,Scherbl2019,Liu.2023}.
To observe them, we turn to the CSD of two such YSR states, akin to those explored in Refs.~\cite{EstradaSaldaa2020,Krtssy2021}.
In \cref{fig:2}b, we show the zero-bias conductance matrix measured as a function $\VRM$ and $\VLM$ when the ABS is tuned away from its energy minimum.
All elements of the conductance matrix show prominent resonances arising from the two charge transitions of each QD. 
The type of avoided crossings observed in the CSD serves as an indication of the coupled-YSR system's ground state \cite{Tsintzis.2022,Dvir.2023}.
Avoided crossings along a negative diagonal, as seen in all four resonances in \cref{fig:2}b, show hybridization between states with the same total charge.
This is the ground state of the system when $t>\Delta$. 
This observation is further confirmed by the negative non-local conductance that is characteristic of ECT \cite{Dvir.2023}. 

Increasing the value of $\VH$ leads to a change in the CSD, shown in \cref{fig:2}c. 
Here, the bottom-left and the top-right resonances show avoided crossings along a positive diagonal associated with $\Delta>t$. 
In the top-left and bottom-right, the resonance lines cross each other, indicating a PMM sweet spot, with $t\approx \Delta$. 
Finally, bringing the ABS close to its energy minimum by a further increase of $\VH$ tunes all of the quadrants of the CSD to the $\Delta>t$ regime (Fig.~\ref{fig:2}d). 
The theoretical model reproduces the observed evolution of the CSDs (\cref{ED:th_coupledYSRs}a--c).

\section{Gate control of CAR and ECT}

\begin{figure*}[ht!]
    \centering
    \includegraphics[width=\textwidth]{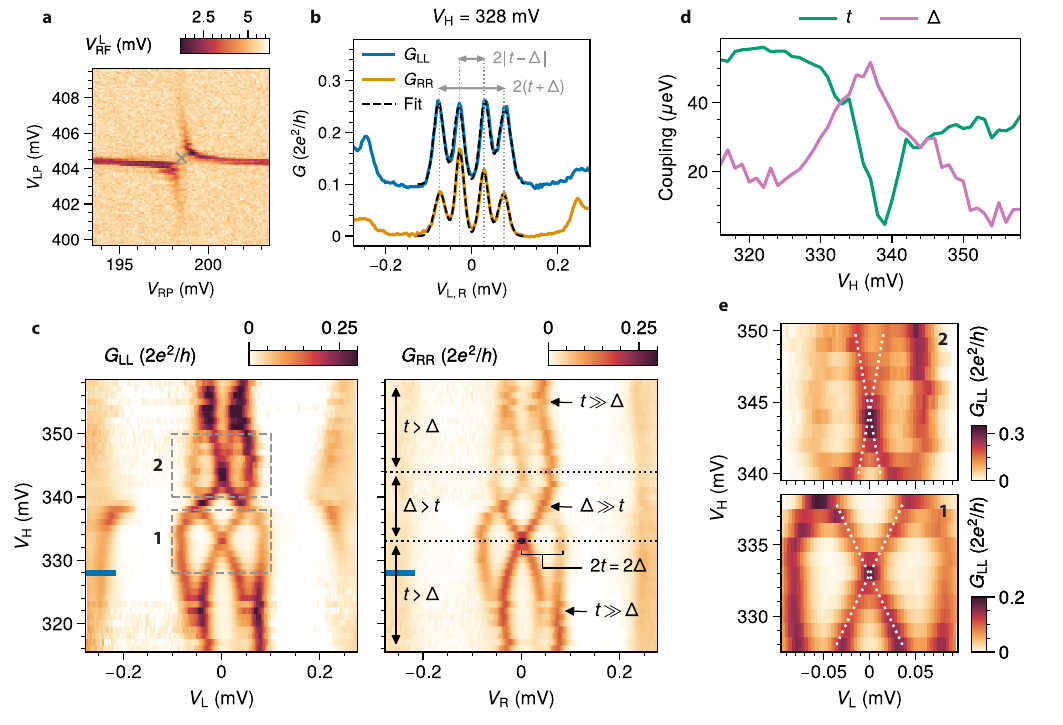}
    \caption{\textbf{ECT and CAR coupling as a function of the ABS chemical potential.} 
    \textbf{a,} CSD measured using RF reflectometry. The center of the avoided crossing, marked with a gray cross, is identified as $\muL=\muR=0$.
    \textbf{b,} Spectrum measured on both sides of the device at the center of the avoided crossing shown in a. The inner peaks correspond to excited states at $\abs{t-\Delta}$, while the outer ones correspond to excited states at $t+\Delta$. By summing and subtracting these energies, it is possible to extract $t$ and $\Delta$ for each $\VH$ value. 
    \textbf{c,} Spectrum as a function of $\VH$ while $\muL=\muR=0$. Each line of the spectrum was measured as described above. The blue tick indicates the $\VH$ value at which a and b were measured.
    \textbf{d,} ECT and CAR couplings extracted from c.
    \textbf{e,} Zoom-in on the zero-energy crossings highlighted in c. The white dotted lines are fits to extract the slope of the linear splitting. 
    The data reported here was collected for the bottom-right charge transition shown in \cref{fig:2}. More details about data processing and data for all charge degeneracies are reported in \cref{ED7:couplings_full}.
    }
    \label{fig:3}
\end{figure*}

ABS-mediated ECT and CAR couplings between QDs are controlled by the chemical potential of ABS \cite{Liu.2022, Bordin.2022}. 
At the energy minimum of the ABS, its excitation is equally electron- and hole-like, both parts interfering constructively to enhance CAR and destructively to quench ECT. 
Finite values of $\muH$ lead to an imbalance between the electron and hole parts of the ABS, decreasing the value of the CAR coupling while enhancing ECT. 
This control over the ECT and CAR amplitudes guarantees a PMM sweet spot when the two QDs are coupled via a single ABS \cite{Liu.2022,Bordin.2022,Tsintzis.2022}. 
To demonstrate that this description can be extended to YSR states coupled via an ABS, we study how the couplings $t$ and $\Delta$ vary as the electrochemical potential of the ABS changes. 

The magnitudes of $t$ and $\Delta$ can be extracted by measuring the excitation spectrum of the system at $\muL=\muR=0$, when spectral splitting is determined by the couplings alone. 
We limit the discussion here to the bottom-right crossing of \cref{fig:2}, noting that the other crossings exhibit qualitatively similar behavior (\cref{ED7:couplings_full}).
At each fixed value of $\VH$, we measure a CSD (\cref{fig:3}a) and set the QD gates to the center of an avoided crossing.
There, the subgap spectrum exhibits two sets of electron-hole symmetric peaks at energies $\abs{t-\Delta}$ and $t+\Delta$ (\cref{fig:3}b), as detailed in Methods. 
By fitting the measured spectrum with two pairs of Gaussians symmetric around $V_\mathrm{L,R}=0$, we extract the energy of the excited states and calculate $t$ and $\Delta$.
We repeat this procedure for different values of $\VH$ and collect the spectra in \cref{fig:3}c, where each line was measured as described above. 

\cref{fig:3}d shows the extracted values of ECT and CAR amplitudes.
At $\VH \approx \SI{322}{mV}$ and $\VH \approx \SI{355}{mV}$, the outer peaks almost merge with the inner ones, indicating $\abs{t-\Delta} \approx t +\Delta$.
Inspecting the corresponding CSDs (see \cref{fig:2}b for an example, all the CSDs are available in the online repository), we can see that ECT is dominant and, therefore, $t\gg \Delta$.
Upon varying $\VH$, the two peaks split into four well-separated peaks, signaling that CAR is increasing although ECT still prevails.
At $\VH \approx \SI{333}{mV}$ and \SI{345}{mV}, the two inner peaks merge into a single zero-bias peak.
These are two sweet spots where $t \approx \Delta$.
In between them, CAR dominates, as the CSDs can confirm (see \cref{fig:2}d).
Around $\VH \approx \SI{338}{mV}$, the outer peaks merge again, this time indicating $\Delta \gg t$. 
This feature is the peak in $\Delta$ and dip in $t$ seen in \cref{fig:3}d, because of the aforementioned interference effects \cite{Liu.2022,Bordin.2022}.
Finally, \cref{ED:th_coupledYSRs}d shows that the same spectral features are reproduced by the theoretical model.

The zero-energy crossings in \cref{fig:3}e allow us to characterize the robustness of the PMM sweet spots to charge fluctuations affecting the ABS and, consequently, $t$ and $\Delta$, causing a splitting of the zero-energy states.
Indeed, we observe that each zero energy state splits with a linear dependence on $\VH$ as predicted by theory, indicating the lack of protection against deviations from the condition $t=\Delta$ known for a two-site Kitaev chain~\cite{Leijnse.2012}.
We extract a slope of $\frac{\partial E}{\partial \VH} \approx \SI{7}{\micro eV/mV}$ and $\SI{3}{\micro eV/mV}$ for the two crossings, comparable to what we measured in Ref.~\cite{Dvir.2023}. 

\section{Majorana sweet spot}

\begin{figure*}[ht!]
    \centering
    \includegraphics[width=\textwidth]{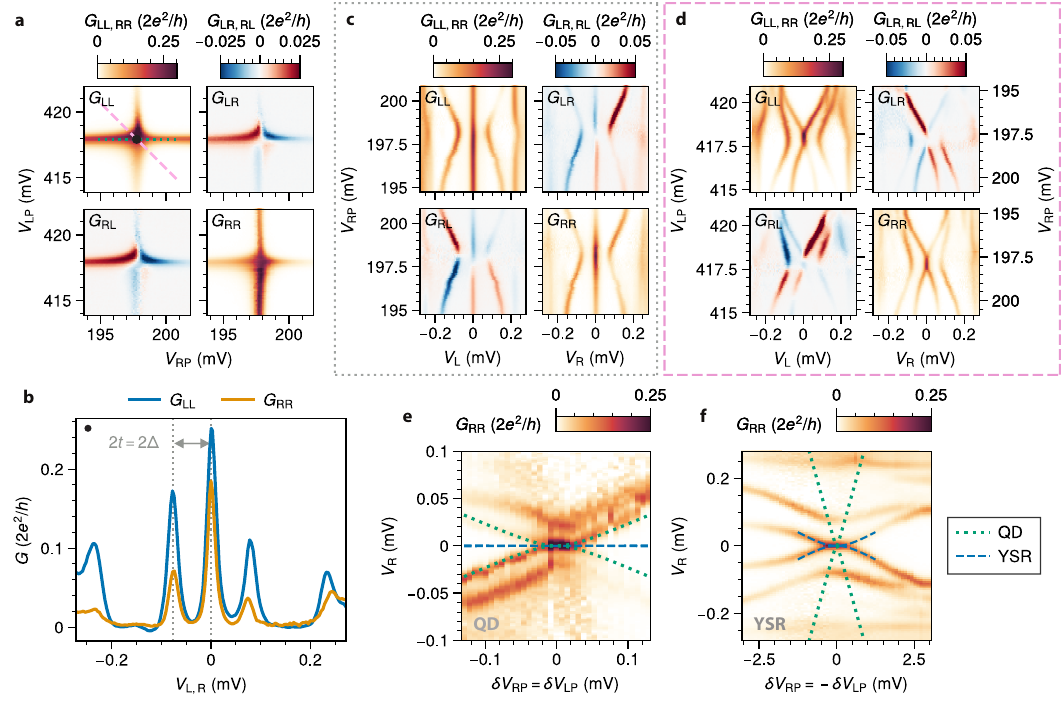}
    \caption{\textbf{Poor man's Majorana sweet spot for coupled YSR states.}
    \textbf{a,} Conductance matrix of a crossing in the CSD, when $t\approx \Delta$.
    \textbf{b,} Spectrum of both sides measured at the sweet spot in the center of the crossing of panel a.
    \textbf{c,} Conductance matrix of the spectrum as a function of $\VRM$ (along the gray dotted line in a). The right QD is detuned across the sweet spot, while the left one is kept on resonance.
    \textbf{d,} Conductance matrix of the spectrum measured detuning both QDs simultaneously along the antidiagonal along the pink dashed line in a.
    \textbf{e, f,} $G_\mathrm{RR}$ as a function of the simultaneous detunings of both sites, $\delta \VLM$ and $\delta \VRM$, away from the sweet spot, using QDs (e, replotted from Ref.~\cite{Dvir.2023}) and YSR states (f, from the same dataset presented in d). For comparison, we have plotted the two gate dispersions with the same scale in both plots. The green dotted lines correspond to the expected energy splitting of a PMM realized with QDs, the blue dashed lines to that of a PMM realized with YSR states, as detailed in the Methods section.
    }
    \label{fig:4}
\end{figure*}

Finally, we measure the spectrum and stability of the PMM states against perturbation of the QDs.
We tune our device to the sweet spot at $\VH=\SI{333.5}{mV}$, where $t\approx\Delta$.
In \cref{fig:4}a, we show the full conductance matrix of the CSD at the sweet spot.
As expected, the resonance lines cross each other and non-local conductance alternates between positive and negative values.
The spectrum measured at the sweet spot in the center of the CSD (\cref{fig:4}b) shows, on both sides of the device, a zero-bias conductance peak clearly separated from the excited states.
As a result of the much stronger coupling between YSR states, the first excited states reside at energies of $\Egap\approx\SI{76}{\micro eV}$, three times larger than in our previous report~\cite{Dvir.2023} and significantly above the electron temperature.

One of the hallmarks of PMM states is their stability against local perturbations.
In \cref{fig:4}c, we measure the spectrum of the QDs varying $\VRM$ while keeping the left QD on resonance.
The observed zero-bias conductance peak persists within the range of the investigated charge degeneracy.
Tuning $\VLM$ while keeping the right QD on resonance (\cref{ED:PMM_extrapaths}b) shows the same qualitative features. 
The same behavior can be reproducibly observed with other QD and ABS orbitals, as shown in \cref{ED:PMM2}. 

If the electrochemical potentials of both QDs are detuned from the sweet spot, PMM states are expected to split quadratically.
This is verified in \cref{fig:4}d, where we measure the spectrum while detuning both QDs along the antidiagonal path shown in \cref{fig:4}a.
The spectrum taken along the diagonal path is shown in \cref{ED:PMM_extrapaths}c. Numerical simulations reproducing these measurements are reported in \cref{ED:th_coupledYSRs}e--i. 
For comparison, we plot the quadratic dispersion measurements of the QD-based vs YSR-based PMMs side by side in panels e and f.
The curvature of the energy-gate dispersion close to the sweet spot is directly proportional to the dephasing rate resulting from charge noise affecting both QDs~\cite{Aasen.2016,Knapp.2018, Mishmash.2020, Boross.2022} and is therefore a measure of the Majorana states' robustness against it.
Comparing the overlaid curves in \cref{fig:4}e and f, we find the gate dispersion curvature reported in this work to be a factor of $\sim 150$ lower than the non-proximitized case (details in Materials and Methods).
This striking reduction can be fully explained by the decreased lever arm of the YSR states, $\alpha$ and the increased $\Egap$, since the curvature expected from the theory model is $\alpha^2/\Egap$ (see Materials and Methods).

To illustrate the effect of reduced charge dispersion on the coherence of a potential poor man's Majorana qubit, we calculate the dephasing rates using realistic charge noise estimations and data presented above (see Supplementary Information).
While perturbation of QDs' potentials is expected to be the dominant mechanism of energy splitting and thus dephasing for non-proximitized PMMs, the drastic reduction in gate dispersion of YSR-based PMMs makes it negligible compared to dephasing caused by deviations from the $t=\Delta$ condition.
The estimated $T_2^* \sim \SI{10}{\nano s}$ is now limited by noise affecting the couplings $t$ and $\Delta$, nearly an order of magnitude higher than the charge-noise-limited $T_2^*$ extracted from data in our previous report~\cite{Dvir.2023}. 
Importantly, the expected dephasing time is now also much longer than the adiabatic limit $\hbar/\Egap \sim \SI{10}{ps}$, which sets an upper bound on how fast Majorana states can be manipulated without populating the excited states \cite{Aasen.2016,Boross.2023}.

\section{Conclusion}

In conclusion, we have demonstrated the formation of YSR states by hybridizing QDs with a common ABS.
The coupling between these YSR states can be controlled by varying the electrochemical potential of the ABS, thus realizing a fully tunable two-site Kitaev chain.
The resulting PMM states have two significant improvements over those in a non-proximitized QD chain.
First, the stronger coupling between the YSR states triples the gap between the PMM and excited states, protecting the former from finite temperature excitations \cite{Knapp.2018} and enabling faster adiabatic operations \cite{Aasen.2016,Boross.2023}.
Second, the reduced charge dispersion of the YSR states enhances the robustness of the PMMs against charge noise affecting both QDs by more than 100.
Thanks to these, even a two-site Kitaev chain realized using YSR states should suffice for a prototypical Majorana qubit and verification of non-abelian properties with fusion and braiding experiments.
Despite the present lack of protection against tunnel-coupling noise, the expected coherence of Majorana qubits made from PMM states is increased by close to an order of magnitude compared to the first report~\cite{Dvir.2023}.
In the future, increasing the number of sites can mitigate noise affecting the tunnel-coupling rates. 
Estimations using parameters of the YSR-based PMMs suggest that a Majorana qubit realized with Kitaev chains as short as 3 to 5 sites could already achieve dephasing times comparable to those predicted for continuous nanowires \cite{Sau.2012,Knapp.2018}.

\section{Data availability and code availability}
Raw data presented in this work, the data processing/plotting code, and code used for the theory calculations are available at \url{https://zenodo.org/records/10013728}.

\section{Acknowledgements}
This work has been supported by the Dutch Organization for Scientific Research (NWO) and Microsoft Corporation Station Q. We thank John M. Hornibrook and David
J. Reilly for providing the frequency multiplexing chips. We thank Martin Leijnse, Ville Maisi, Pasquale Scarlino, Athanasios Tsintzis, Rubén Seoane Souto, Karsten Flensberg, Srijit Goswami, Sebastiaan L.\ D.\ ten Haaf, Qingzhen Wang, Ivan Kulesh, and Yining Zhang for helpful discussions. 

\section{Author contributions}
DvD, FZ, AB, GPM, NvL, and JCW fabricated the device. FZ, DvD, DX, BR, and TD performed the electrical measurements. TD and GW designed the experiment. FZ, DvD, GW, and TD analyzed the data. FZ, DvD, TD, GW, and LPK prepared the manuscript with input from all authors. TD and LPK supervised the project. CXL and AMB developed the theoretical model with input from MW. G.B., S.G., and E.P.A.M.B. performed InSb nanowire growth.

\section{Competing interests}
The authors declare no competing interests.

\bibliography{bibliography}

\pagebreak

\section{Materials and Methods}

\subsection{Device fabrication}
The InSb/Al hybrid nanowire device presented in this work was fabricated using the shadow-wall lithography technique \cite{Heedt.2021,Borsoi.2021}. 

A substrate is patterned with Ti/Pd gates. $\SI{10}{\nano m}$ of AlO$_x$ and $\SI{10}{\nano m}$ of HfO$_x$ are deposited by ALD as gate dielectric. HSQ shadow-walls are then patterned.
Nanowires are deposited and pushed next to the shadow-walls using an optical micro-manipulator. An $\SI{8}{\nano m}$ Al shell is deposited at alternating angles of $\SI{15}{\degree}$ and $\SI{45}{\degree}$ with respect to the substrate, followed by a capping layer of $\SI{20}{\nano m}$ of AlO$_{x}$. Normal leads in ohmic contact with the nanowire are fabricated by Ar milling and evaporation of Cr/Au. 

Additional details about the substrate fabrication and the Al deposition are described in Ref.~\cite{Mazur.2022}.

\subsection{Transport measurements and data processing}

The measurements are done in a dilution refrigerator with a base temperature of $\SI{30}{\milli K}$. A magnetic field of $\SI{150}{\milli T}$ is applied along the nanowire axis except in \cref{ED3:QDs_sweetspot}c,d.

The three-terminal setup used to measure the device is illustrated in \cref{fig:1}a. The SC lead is always kept grounded.
The two normal leads can be voltage-biased independently.
When a bias is applied to one side, the other one is kept grounded.
The currents $I_L$ and $I_R$ are measured separately.
Digital multimeters and lock-in amplifiers are used to read the voltage outputs of the current meters.
AC excitations of $\SI{5}{\micro V}$ RMS are applied on each side with different frequencies ($\SI{39}{Hz}$ on the left and $\SI{29}{Hz}$ on the right), except for \cref{ED1:hybrid_spectrum} and \cref{ED11:QDABS_right}d--f, where excitations of $\SI{10}{\micro V}$ RMS were used.

Off-chip multiplexed resonators~\cite{Hornibrook.2014} connected to the two normal leads are used for fast RF reflectometry measurements.
This measurement scheme was employed to speed up the tune-up of the device, for \cref{fig:3}, and for \cref{ED7:couplings_full}, as explained in \cref{fig:3}a.
Additional details about the reflectometry setup are described in Ref.~\cite{wang2022parametric}.

\subsection{Spinless PMM spectrum}

The low-energy spectrum of the coupled YSR system arising from the model described in Supplementary Information can be reduced to that of a spinless PMM model \cite{Leijnse.2012,Liu.2022,Tsintzis.2022,Liu.2023,Tsintzis.2023}. 

In this simplified model, the non negative energy eigenvalues are \cite{Leijnse.2012}
\begin{equation*}
    E_{\pm} = \sqrt{\mu_+^2 + \mu_-^2+t^2+\Delta^2 \pm 2\sqrt{(\mu_+^2+\Delta^2)(\mu_-^2+t^2)}},
\end{equation*}
where $\mu_\pm = \frac{\muL \pm \muR}{2}$, $\muL$ and $\muR$ are the chemical potentials of the two sites, and $t$, $\Delta$ are the couplings. The spectrum is symmetric around zero energy because of particle-hole symmetry. If $\muL=\muR=0$, the eigenvalues reduce to $E_+ = t+\Delta$ and $E_- = \abs{t-\Delta}$. These are the excitation energies that we use to extract $t$ and $\Delta$ from the measured spectra in \cref{fig:3}.

The gate dispersion lines plotted in \cref{fig:4}e,f are calculated using the lowest excitation energy. When $t=\Delta$ we obtain
\begin{equation*}
    E_- = \sqrt{\alpha^2 \delta V^2 + 2t^2-2\sqrt{t^2(t^2+\alpha^2 \delta V^2)}},
\end{equation*}
where $\alpha$ is the lever arm converting voltage to chemical potential and $\delta V$ is the simultaneous detuning of each gate away from the sweet spot. 
For the PMM realized with QDs, we have used $\alpha = 0.33e$ and $t = \SI{12}{\micro eV}$ \cite{Dvir.2023}. For the PMM realized with YSR states we have used $\alpha = 0.05e$ (\cref{ED3:QDs_sweetspot}) and $t = \SI{38}{\micro eV}$ (\cref{fig:4}b).
Finally, the quadratic splitting of PMM states can be derived by expanding the expression above for small $\delta V$, which gives
\begin{equation*}
    E_- \approx \frac{\alpha^2}{2t}\delta V^2.
\end{equation*}

\subsection{Series resistance correction}

The effect of series resistance in the fridge line and other parts of the circuit on transport measurements of a three-terminal device of our type is described in Ref.~\cite{Martinez2021}, as well as how to correct for it.
In our setup, the resistance of the voltage source is $\SI{100}{\Omega}$ and that of the current meter is $\SI{200}{\Omega}$.
Additional series resistance comes from the fridge lines and the ohmic contacts.
Correcting for the voltage fall over these resistors in series to the device requires knowledge of the exact resistance values, which we presently cannot obtain before the next sample exchange.
Therefore, we make use of the fact that the bulk Al superconducting gap  in local conductance measurements should stay constant across all gate values to arrive at an estimation of the total series resistance in each fridge line.

In \cref{ED8:Rseries}a,c it is possible to observe that the superconducting gap appears enlarged when there is finite subgap conductance, indicating unaccounted-for series resistance.
To estimate it, we correct the $\GLL$ measurement shown in \cref{ED8:Rseries}a for different trial values of the series resistance.
We extract the energies of the coherence peaks and calculate their variance.
Finally, we choose the value that minimizes the variance as the optimal series resistance. 
We find an optimal value of $\SI{3.65}{k\Omega}$, in addition to the resistance of the voltage source and current meter.
The corrected measurement is shown in \cref{ED8:Rseries}b.
All the data processing steps are available in the online repository.
In \cref{ED8:Rseries}c,d we show that the same series resistance value also corrects an analogous measurement on the right side.
In the regimes relevant to this work, the voltage divider effect caused by the series resistance does not strongly affect local conductance.
On the other hand, it can be more pronounced for non-local conductance, as discussed in \cref{ED8:Rseries}e,f.
Using the extracted resistance value, we apply the correction method described in Ref.~\cite{Martinez2021} to the data presented in \cref{fig:2}, \cref{fig:3}, \cref{fig:4}, \cref{ED1:hybrid_spectrum}, 
\cref{ED11:QDABS_right}d, \cref{ED7:couplings_full}, \cref{ED:PMM_extrapaths}, \cref{ED:PMM2}d--k, and \cref{ED8:Rseries}b,d,f, where the full conductance matrix is measured.

\subsection{Device tune-up}

The device is controlled with seven bottom gates (\cref{fig:1}a).
To perform tunnel spectroscopy of the hybrid segment in \cref{ED1:hybrid_spectrum} and \cref{ED11:QDABS_right}, we use $\VLI$ and $\VRI$ to form a tunnel barrier and apply a large positive voltage to $\VLM$, $\VRM$, and the outer gates. 
$\VH$ is used to control the electrochemical potential in the hybrid segment, where discrete states are confined because of the tunnel barriers.
QDs are formed by reducing the voltage on the outer gates.
This creates additional tunnel barriers next to the normal leads, thus shaping a confining potential.
$\VLM$ and $\VRM$ control the electrochemical potential of the QDs.
Finally, the coupling between the QDs and the ABS can be controlled using $\VLI$ and $\VRI$.
By increasing their voltage, we can achieve the strong coupling regime between the QDs and the ABS in the hybrid as explained in \cref{fig:1}.
For all the measurements presented in the main text, the gates defining the tunnel barriers were maintained at the same value.

\pagebreak

\section{Supplementary Information}
\subsection{Theoretical model}
The physical system of double QDs connected by a hybrid segment in the middle can be well described by a three-site model as below 
\begin{align}
& H = \sum_{i=L,H,R} H_i + H_{T}, \nn
& H_i = (\mu_i + E_{Zi}) n_{i\su} +  (\mu_i - E_{Zi}) n_{i\sd} + U_i n_{i\su} n_{i\sd} +  \Gamma_i (c\dg_{i\su} c\dg_{i\sd} +  c_{i\sd} c_{i\su} ), \nn
& H_T = w_L ( c\dg_{H\su} c_{L\su}  +  c\dg_{H\sd} c_{L\sd}  ) + w_{SOL} ( c\dg_{H\sd} c_{L\su}  -  c\dg_{H\su} c_{L\sd}  ) \nn
& \quad + w_R ( c\dg_{R\su} c_{H\su}  +  c\dg_{R\sd} c_{H\sd}  ) + w_{SOR} ( c\dg_{R\sd} c_{H\su}  -  c\dg_{R\su} c_{H\sd}  ) + h.c..
\label{eq:ham_three_site}
\end{align}
Here $H_i$ is the Hamiltonian for the QDs ($i=L,R$) or the ABS ($i=H$) in the hybrid, $n_{i\sigma} = c\dg_{i\sigma}c_{i\sigma}$ is the occupancy number of the QD orbital or normal state in the hybrid with spin $\sigma$, $\mu_i$ is the chemical potential energy,  $E_{Zi}$ is the induced Zeeman energy, $U_i$ is the charging energy, and $\Gamma_i$ is the induced superconducting gap.
$H_T$ is the tunnel Hamiltonian between the QDs and the ABS, where $w_{L/R}$ and $w_{SOL/SOR}$ are the amplitudes for the spin-conserving and spin-flipping single electron tunneling processes.
The parameters used for the numerical simulations performed in this work, in unit of $\Gamma_H\equiv \Gamma$, the induced gap on the ABS, are: $E_{ZL}=E_{ZR}=1.5\Gamma$, $E_{ZH}=0.5\Gamma$ for Zeeman energy, $U_{L}=U_{R}=5\Gamma$, $U_{H}=0$ for charging energy, $w_{L}=\Gamma$, $w_{SOL}=w_{L}/2$, $w_{R}=0.75\Gamma$, $w_{SOR}=w_R/2$ for tunneling amplitudes, and $\Gamma_L=0, \Gamma_R=\Gamma$ for induced gaps on the QDs. 
Here the magnitude of the Zeeman energy on the ABS is weaker than that in the QDs owing to the $g$ factor renormalization in the hybrid. Furthermore, an extra pairing gap is added on the right QD to explain the residual Andreev conductance when the ABS is far from its energy minimum, as shown in the measurement in \cref{fig:1}.
The sweet spot investigated in \cref{ED:th_coupledYSRs}e--i corresponds to $\mu_H = 0.617\Gamma$, $\mu_L = 2.290\Gamma$, and $\mu_R = -5.879\Gamma$.
Finally, in \cref{ED:th_QDABS_wf}, we calculate the electron and hole character of an excitation between two states of different parity as
\begin{equation}
    u_i^2 = \sum_{\sigma=\su, \sd} \abs{\bra{o}c_{i\sigma}^\dagger \ket{e}}^2, \quad v_i^2 = \sum_{\sigma=\su, \sd} \abs{\bra{o}c_{i\sigma}\ket{e}}^2,
\end{equation}
where $\ket{e}$ is the even state and $\ket{o}$ the odd one.

In the experimental device, conductance spectroscopy of the hybrid system is measured in a three-terminal junction with two normal leads and one grounded superconducting lead.
To numerically simulate the transport measurement, we attach two additional normal leads to the two QDs, respectively.
The lead Hamiltonians are
\begin{align}
H_{\text{lead}} = \sum_{a=L,R}\sum_{\sigma=\su, \sd}   \sum_{k}  (\varepsilon_{ka} - \mu_{\text{lead},a}) f\dg_{ak\sigma} f_{ak\sigma},
\end{align}
where $f_{ak\sigma}$ is the annihilation operator of a free electron with momentum $k$ and spin $\sigma$ in lead-$a$, $\varepsilon_{ka}$ is the kinetic energy, and $\mu_{\text{lead}, a}$ is the lead chemical potential.
The tunnel Hamiltonian between leads and the QDs is
\begin{align}
H_{T, \text{lead}} = \sum_{a=L,R}\sum_{\sigma=\su, \sd}   \sum_{k}  w'_a f\dg_{ak\sigma} c_{a\sigma} + h.c.,
\end{align}
where $w'_a$ is the tunneling amplitude between QD-$a$ and lead-$a$, and is assumed to be spin-diagonal and independent of $k$.
A more useful physical quantity to characterize the lead coupling strength is $\gamma_a = 2\pi \rho w'^2_a$ with $\rho$ being the density of states at the Fermi energy of the normal lead.
Based on the model Hamiltonians introduced above, current and conductance matrices can be numerically calculated by solving a rate equation for the reduced density matrix of the hybrid system. 
We set the temperature $k_BT=0.04\Gamma$ and tunneling strength $\gamma_a =0.03\Gamma$ to be the same for both normal leads.
Finally, when simulating the local Andreev conductance of a QD-ABS system, we are actually considering a two-terminal junction by removing the quantum dot and the normal lead on the other side.

\subsection{Estimation of dephasing rates}

A minimal Majorana qubit consists of two subsystems, each hosting one pair of Majoranas: $\gamma_{A1},\gamma_{A2}$, and $\gamma_{B1},\gamma_{B2}$.
It is possible to encode a qubit in the subspace of either even or odd global fermion parity.
Without loss of generality, we focus on the even subspace. 
We define qubit states $\ket{0}\equiv \ket{o_A,o_B}$, with the fermion parity of each subsystem being odd, and $\ket{1}\equiv\ket{e_A,e_B}$, each being even.
The qubit Pauli operator $\sigma_z$ is then identified as $i\gamma_{A1}\gamma_{A2}=i\gamma_{B1}\gamma_{B2}$ and the effective Hamiltonian is \cite{Aasen.2016,Boross.2022,Tsintzis.2023}
\begin{equation}
H = \frac{i}{2}\left(E_A\gamma_{A1}\gamma_{A2}+E_B\gamma_{B1}\gamma_{B2}\right)= E \sigma_z,
\end{equation}
where $E_{A,B}=E_{o_{A,B}}-E_{e_{A,B}}$ is the excitation energy of each subsystem and $E\equiv (E_A+E_B)/2$.

An ideal Majorana qubit has $E=E_A=E_B=0$.
Uncontrolled fluctuations of $E$ lead to qubit dephasing \cite{Aasen.2016,Knapp.2018,Mishmash.2020,Boross.2022}.
If the typical amplitude of $E$ fluctuation is $\delta E$, the dephasing rate is $1/ T_2^* \sim \delta E/\hbar$ \cite{Koch.2007,Aasen.2016}. 
In the following estimations, we assume the two subsystems are similar and omit details of the statistical distribution of the noise, so that $\delta E_{A} \sim \delta E_{B} \sim \delta E$.
Thus, we use the typical energy fluctuations of one pair of PMM to estimate the dephasing rates of a qubit.

Energy splitting of a pair of PMM can occur due to two reasons: the degeneracy splits quadratically when the QDs are detuned simultaneously ($\muL = \pm \muR \neq 0$) and linearly when the ECT and CAR couplings are perturbed from the sweet spot ($t \neq \Delta$)~\cite{Leijnse.2012}.
Other types of noise, e.g., magnetic fluctuations, enter the Hamiltonian parametrically via these two mechanisms.
In the main text, we have characterized the sensitivity of the degeneracy to each of them by intentionally tuning gate voltages away from the sweet spots.
These gate dispersion results allow us to estimate the dephasing rate given typical noise amplitudes found in QD qubit devices.
Concretely, we consider random fluctuations of the gate voltages controlling the electrochemical potential of the QDs ($\VLM$, $\VRM$) and ABS ($\VH$), the former causing on-site potential fluctuation and the latter that of the tunneling amplitudes.

We first focus on charge noise affecting the QDs.
In \cref{fig:4}f, we have measured the quadratic splitting of the PMM when the electrochemical potential of both QDs is detuned from the sweet spot.
Given the vanishing first derivative of the gate dipersion, we extracted a second derivative of $\partial ^2 E/\partial \VLM ^2 \approx \SI{30}{e/V}$, two orders of magnitude smaller than the one measured in our previous report, $\partial ^2 E/\partial \VLM ^2 \approx \SI{4500}{e/V}$ \cite{Dvir.2023}.
In previous measurements of charge fluctuations of a QD \cite{Petersson.2010, Dial.2013,Scarlino.2022,Connors.2022,Burkard.2023}, equivalent voltage noise on the gates is typically found to be $\delta V \sim \SI{10}{\micro V}$ (after accounting for the gate lever arm).
A worst-case estimation of maximally (anti)correlated gate noise on both QDs predicts a resulting energy fluctuation of $\delta E \approx \frac{\partial ^2 E}{\partial \VLM ^2}\delta V^2$, yielding $\sim \SI{3}{neV}$ for the YSR-based PMM states in this work and $\sim \SI{450}{neV}$ for our previous work.
These correspond to dephasing rates of $\delta E/\hbar \sim \SI{5}{MHz}$ and $\sim \SI{700}{MHz}$ for the two cases, a direct consequence of the two orders of magnitude of gate dispersion suppression.

We now turn to charge noise affecting the ABS and thereby the couplings between QDs, $t$ and $\Delta$.
Given the charge sensitivity of the hybrid ABS and its key effect on modulating the tunneling amplitudes, we assume the primary source of $t-\Delta$ variation to be charge fluctuations of the ABS.
In \cref{fig:3}e, we have estimated the noise sensitivity of two PMM sweet spots against perturbation on $\VH$.
We found $\partial E/\partial \VH \approx 3$ to \SI{7}{\micro eV/mV}, comparable to our previous work \cite{Dvir.2023}.
Although the exact sensitivity depends on microscopic details, we note that the superconducting lead, in general, strongly screens charge fluctuations.
This explains the small lever arm for an ABS in the hybrid segment, which is $\sim 1/20$ that of a normal QD (see \cref{ED1:hybrid_spectrum} and \cref{ED2:QDs_highbias}a).
By considering the worst sensitivity and the same fluctuations as before, we estimate typical splittings of $\delta E \sim \frac{\partial E}{\partial \VH} \delta V = \SI{70}{neV}$ and a dephasing rate of $ \sim \SI{100}{MHz}$. 

Comparing the dephasing rates, we find that on-site QD charge fluctuations likely dominated dephasing of PMM qubits made from non-proximitzed QD states.
With the introduction of YSR-based Kitaev chain and its strongly suppressed gate dispersion, the QD charge fluctuation will likely no longer be the main source of dephasing.
Instead, tunneling amplitude variations, not suppressed by the formation of YSR states, becomes the more probable primary dephasing mechanism.
While reducing sensitivity to tunneling amplitude fluctuations is beyond the scope of this paper, we remark here that a possible solution is to increase the number of sites of the Kitaev chain to at least 3~\cite{Sau.2012}. 

Finally, we comment on a few additional details not addressed by the discussions above.
We estimated energy fluctuations only using equivalent gate fluctuations converted from previous charge noise measurements.
This does not describe other sources of decoherence including that due to quasiparticles \cite{Rainis.2012,Knapp.2018} and electron-phonon coupling \cite{Knapp.2018}.
We do not expect our estimations to be drastically altered if these effects are included, given the low electron temperature and that electron-phonon dephasing rates measured in charge qubits realized in III/V semiconductors are on the order of $\SI{100}{MHz}$ \cite{Hofmann.2020,Ranni.2023}, comparable to our estimated dephasing rates due to tunneling-rate variations.
Ultimately, a realistic quantitative noise model would require the measurement of the noise spectrum and dephasing times. 
Despite all these simplifications, the dephasing rate estimations indicate that there should be a reasonable time window where it is possible to adiabatically manipulate PMM states before they decohere significantly.
This is a necessary condition to demonstrate qubit, fusion, and braiding protocols based on PMM states  \cite{Liu.2022.Fusion,Boross.2023, Tsintzis.2023}.
In the future, longer Kitaev chains are expected to increase the dephasing time, since the energy splitting is predicted to decrease exponentially with the number of sites \cite{Sau.2012}.

\pagebreak

\section{Extended Data}

\setcounter{figure}{0}
\renewcommand{\thefigure}{ED\arabic{figure}}
\renewcommand{\theHfigure}{ED\arabic{figure}}

\begin{figure*}[ht!]
    \centering
    \includegraphics[width=0.66\textwidth]{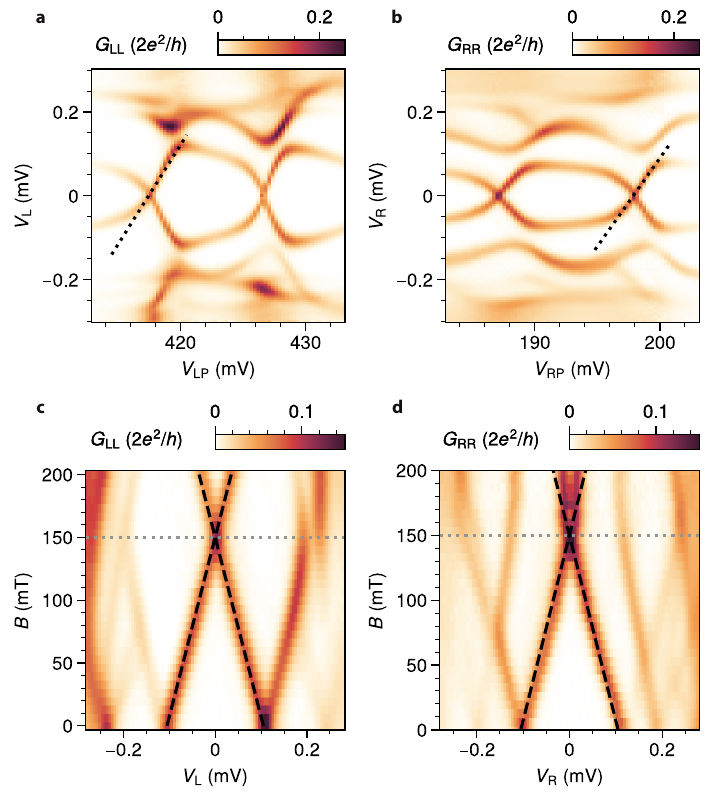}
    \caption{\textbf{Characterization of the YSR states at $t\approx\Delta$.} \textbf{a,} Local conductance measurement of the spectrum of the left YSR state when the right one is off-resonance. $\VH = \SI{333.5}{\milli V}$ is tuned such that $t\approx \Delta$ for the crossing showed in \cref{fig:4}a. The lever arm at charge degeneracy is $\approx 0.05e$, extracted using the black dotted line.
    \textbf{b,} Same as panel a, but for the right YSR state. We extract a lever arm of $\approx 0.04e$.
    \textbf{c,}  Local conductance measurement of the Zeeman splitting of the left YSR state, at the same value of $\VH$ as above. The field is applied along the nanowire axis. The gray dotted line indicates the field at which all the other measurements were conducted.
    The left dot was positioned at the charge degeneracy point at $B = \SI{150}{\milli T}$ and the right one off-resonance. We extract a g-factor of $\approx 24$, consistent with previously measured values for strongly proximitized states in similar devices~\cite{Mazur.2022}. The corresponding Zeeman splitting at $\SI{150}{\milli T}$ is $\approx \SI{200}{\micro eV}$.
    \textbf{d,} Same as panel c, but for the right YSR state. We extract the same g-factor as for the left YSR state.  
   }\label{ED3:QDs_sweetspot}
\end{figure*}

\begin{figure*}[ht!]
    \centering
    \includegraphics[width=\textwidth]{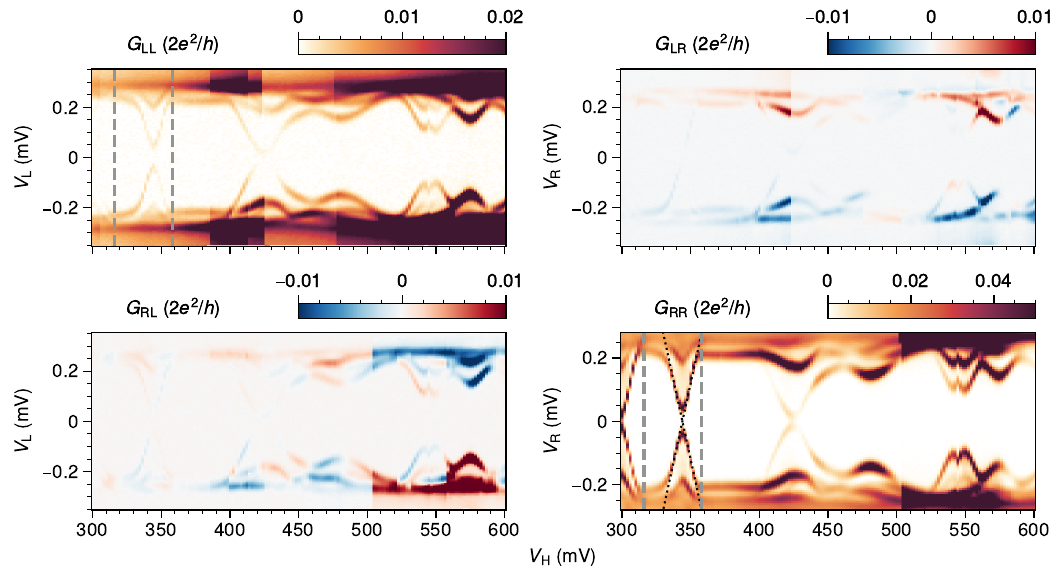}
    \caption{\textbf{Spectroscopy of the hybrid segment.} Conductance matrix of the hybrid spectrum as a function of $\VH$. The hybrid segment hosts separated ABS that can mediate the coupling between the QDs. The gray dashed lines indicate the range of $\VH$ used for the measurements presented in the main text. The lever arm is $\alpha\approx 0.02e$ as extracted from the black dotted lines superimposed on the ABS in panel d, indicating a strong screening effect of the superconducting lead.
    For this measurement, the two sides are tuned as tunnel barriers as described in the Methods section. When QDs are formed, $\VLI$ and $\VRI$ are increased to reach the strong coupling regime between them and the ABS. This affects the ABS spectrum because of cross-coupling and modification to the confining potential.
    Therefore, the $\VH$ values and the energy of the ABS in this measurement are slightly different from those presented in the main text.
   }\label{ED1:hybrid_spectrum}
\end{figure*}

\begin{figure*}[ht!]
    \centering
    \includegraphics[width=\textwidth]{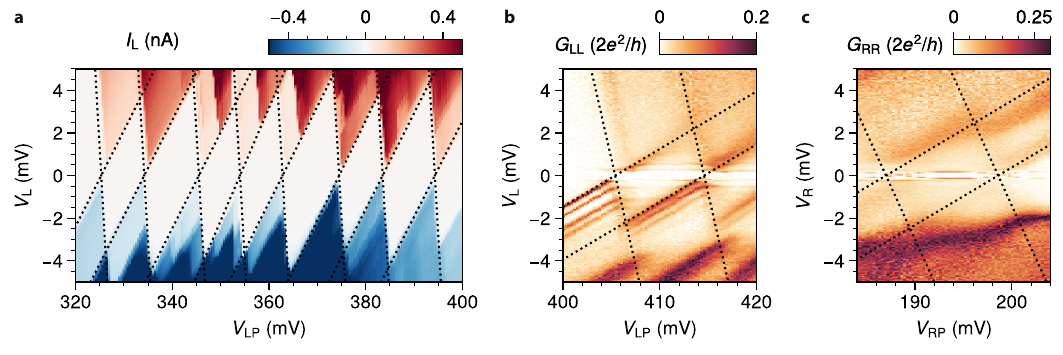}
    \caption{\textbf{High-bias spectroscopy of the QDs.} 
    \textbf{a,} Coulomb diamonds for the left QD when it is not coupled to any ABS by closing $\VLI$ and keeping ABSs off-resonance. From the overlaid black dotted lines, we extract typical charging energies of $\approx \SI{3.5}{\milli eV}$ and lever arms of $\alpha\approx 0.4e$.
    \textbf{b,} Local conductance spectroscopy of the left QD as a function of $\VLM$ after $\VLI$ is increased to strongly couple to an ABS. The measurement was taken in the same configuration as \cref{fig:1}g, with the ABS off-resonance. We extract a charging energy of $\approx \SI{2}{\milli eV}$ and a lever arm of $\alpha\approx 0.2e$, which are renormalized because of the lowered tunnel barriers.
    \textbf{c,} Same as panel b, but for the right QD. Typical QD features are not as clearly visible since the state is significantly proximitized by the SC, as shown in \cref{fig:1}j and in \cref{ED11:QDABS_right}a--c. By overlaying the visible features, we extract a charging energy of $\approx \SI{2}{meV}$ and a lever arm of $\alpha \approx 0.2e$. These measurements were taken applying a field of $\SI{150}{\milli T}$ along the nanowire axis.
   }\label{ED2:QDs_highbias}
\end{figure*}

\begin{figure*}[ht!]
    \centering
    \includegraphics[width=\textwidth]{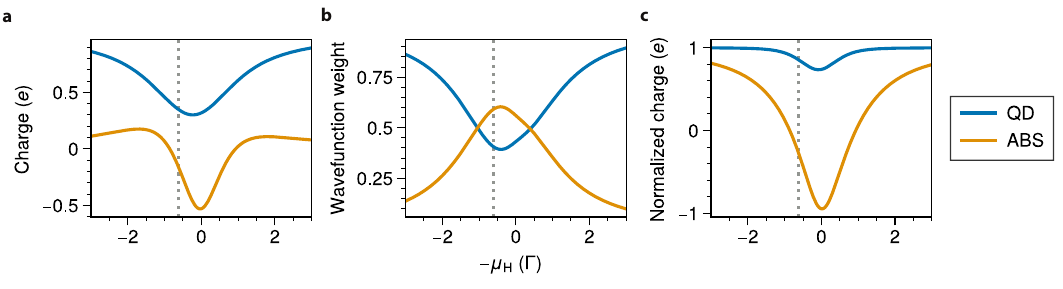}
    \caption{\textbf{Properties of the YSR zero-energy excitation.}
    \textbf{a,} 
    Charge of the YSR excitation residing on the left QD (blue line) and on the ABS (orange line). The charge is calculated as $u_i^2-v_i^2$, with $i=L$ for the QD and $i = M$ for the ABS, as defined in Supplementary Information. The gray dotted line corresponds to the $\muH$ value used for the sweet spot studied in \cref{ED:th_coupledYSRs}e--i. The decrease of the charge residing on the QD explains part of the lever arm reduction observed for YSR states (\cref{fig:1}h). A further reduction comes from charging energy renormalization. 
    \textbf{b,} Weight of the YSR excitation wavefunction on the left QD (blue line) and on the ABS (orange line). The weight of the wavefunction on each site is defined as $u_i^2+v_i^2$. When the ABS is close to its energy minimum at $\muH=0$, the hybridization between the QD and the ABS becomes stronger and the excitation is delocalized on the two sites, thus reducing the charge residing on the QD.
    \textbf{c,} Charge of the YSR excitation normalized for the wavefunction weight residing on the left QD (blue line) and on the ABS (orange line). The normalized charge is defined as  $(u_i^2-v_i^2)/(u_i^2+v_i^2)$ and it quantifies the mixed electron and hole character. It ranges from $1$ to $-1$ when the excitation is purely electron- and hole-like, respectively. 
    Although the charge of a YSR state is not a good quantum number while ECT/CAR is defined using charge states, the predominance of a charge character justifies our use of ECT and CAR in the main text, as investigated in more detail in Ref.~\cite{Liu.2023}.
    }\label{ED:th_QDABS_wf}
\end{figure*}

\begin{figure*}[ht!]
    \centering
    \includegraphics[width=\textwidth]{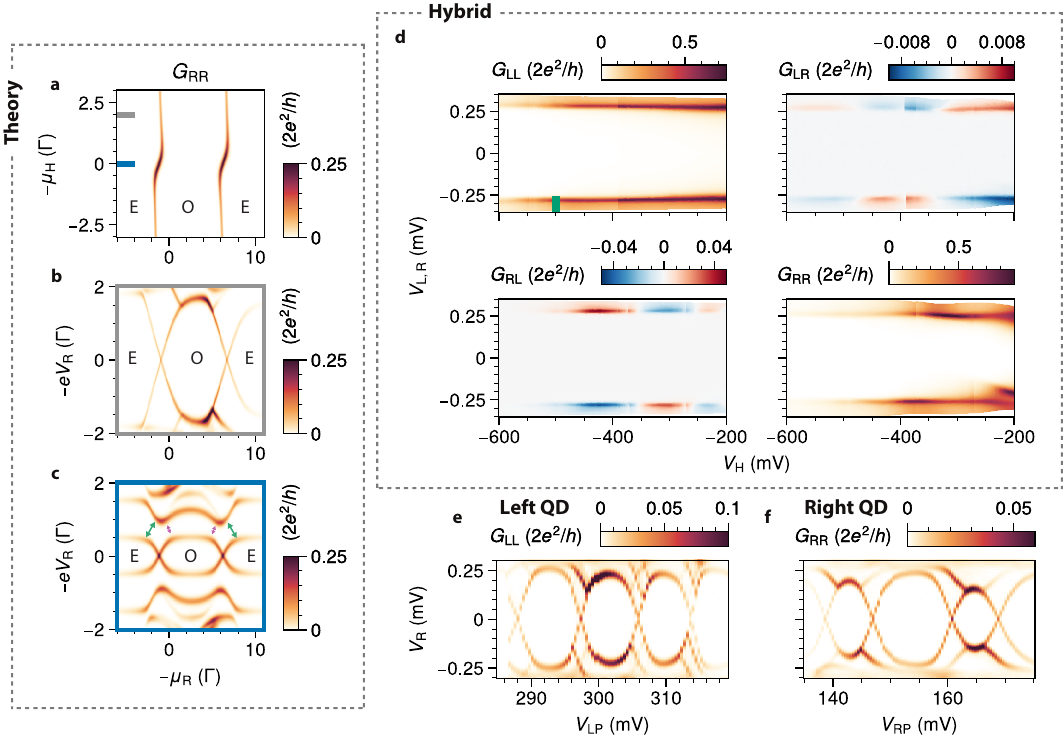}
    \caption{\textbf{QDs directly proximitized by the superconducting film.} 
    \textbf{a--c,} Same as \cref{fig:1}c--e, but with an additional gate-independent source of proximity for the QD, as discussed in Supplementary Information.
    \textbf{d,} Conductance matrix for tunnel spectroscopy of the hybrid section as a function of $\VH$. The green tick marks the value of $\VH$ at which the measurements in e and f were taken. In this regime, there are no subgap states.
    \textbf{e, f,} Local tunnel spectroscopy of the left (\textbf{b}) and right (\textbf{c}) QDs at low $\VH$. Although there are no subgap states in the hybrid, it is possible to tune the QDs in a regime where a gap and subgap states appear. We interpret this as arising from the QDs being proximitized by the Al, indicating that the proximity of the QDs can be not exclusively due to subgap states in the hybrid. These measurements were taken applying a field of $\SI{150}{\milli T}$ along the nanowire axis.
    }\label{ED11:QDABS_right}
\end{figure*}

\begin{figure*}[ht!]
    \vspace{-2cm}
    \centering
    \includegraphics[width=\textwidth]{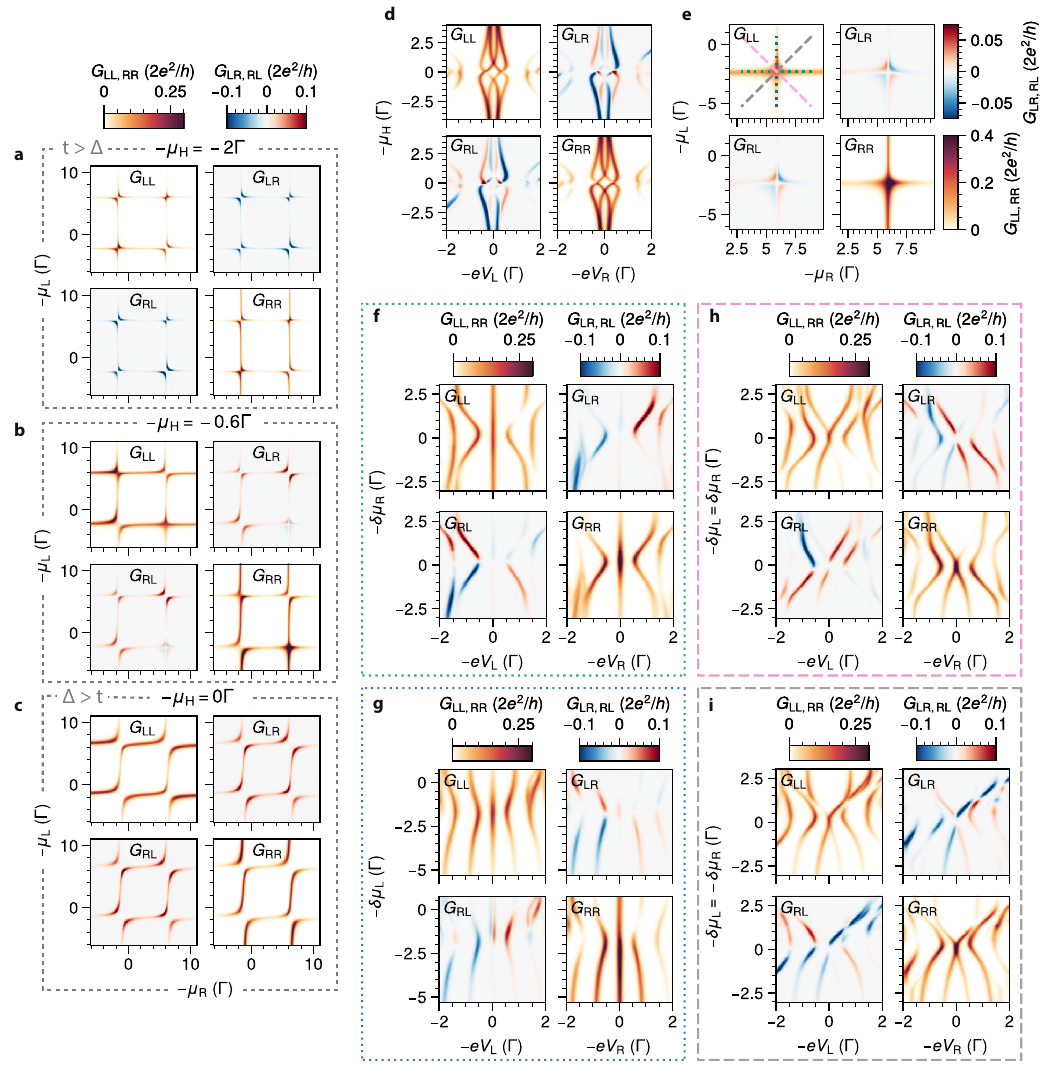}
    \caption{\textbf{Numerical simulations of coupled YSR states}
    \textbf{a-c} Numerical conductance matrix of the CSDs of coupled YSR states varying $\muH$, reproducing \cref{fig:2}.
    \textbf{d,}  Numerical conductance spectrum of coupled YSR states as a function of the ABS chemical potential. The spectrum is calculated following the same approach discussed in \cref{fig:3} to maintain the chemical potentials of the QDs at the center of the crossing. The theoretical model reproduces the experimental findings presented in \cref{fig:3}.
     \textbf{e,} Numerical conductance matrix of a crossing in the CSD, when $t\approx \Delta$, at $\muH = 0.617\Gamma$.
    \textbf{f, g,} Numerical conductance matrix of the spectrum as a function of the right (f)/left (g) QD detuning along the green/blue dotted line in panel a. The other QD is kept on resonance. The numerical results reproduce the measurement shown in \cref{fig:4}c/\cref{ED:PMM_extrapaths}b.
    \textbf{h, i,} Numerical conductance matrix of the spectrum calculated detuning both QDs simultaneously, along the pink/gray dashed line in panel e. The numerical results reproduce the measurements shown in \cref{fig:4}d/\cref{ED:PMM_extrapaths}c.
   }\label{ED:th_coupledYSRs}
\end{figure*}

\begin{figure*}[ht!]
    \centering
    \includegraphics[width=\textwidth]{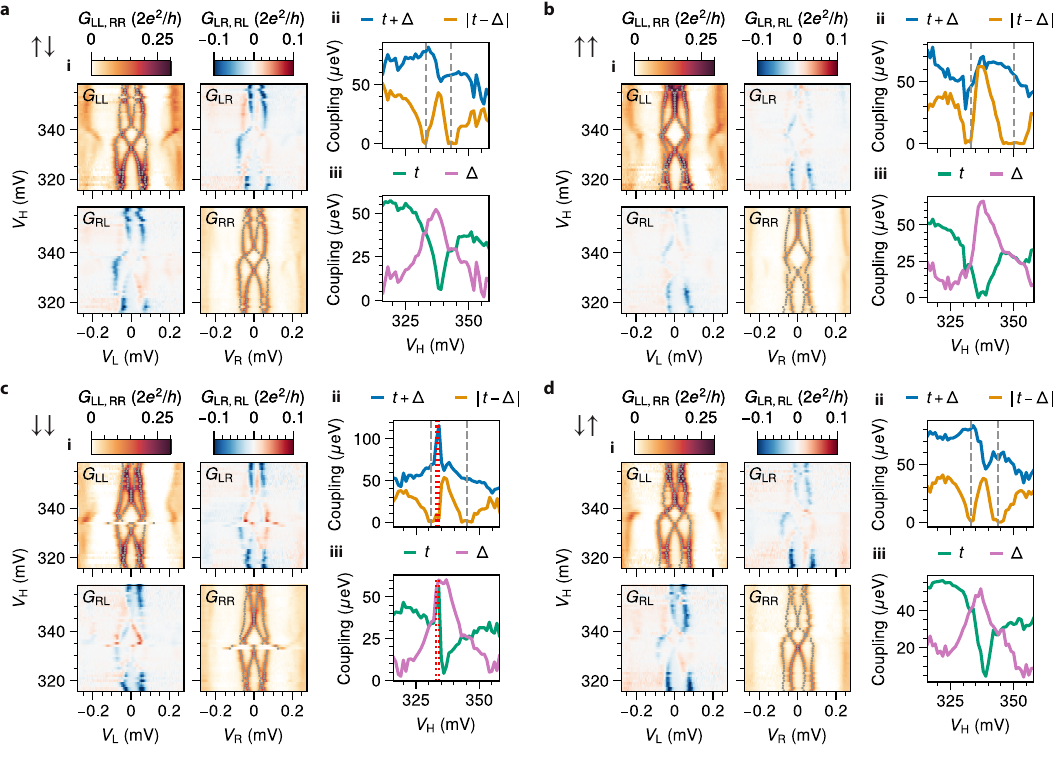}
    \caption{\textbf{ECT and CAR coupling for all the resonances.} 
    \textbf{a--d,} (i) Conductance matrix of the spectrum of two coupled YSR states as a function of $\VH$, measured as described in \cref{fig:3}. The extracted peaks are marked with gray dots. (ii) Energy of the excited states, calculated by averaging the peaks at positive and negative bias of both the left and the right spectrum. From the outer peaks we extract $t+\Delta$, from the inner ones $\abs{t-\Delta}$. The gray dashed lines indicate a change of sign of $t-\Delta$. In particular, $t<\Delta$ within the lines and $t>\Delta$ outside.
    In c a gate jump affected the measurement of the spectrum for two values of $\VH$. These are highlighted with red dotted lines in c(ii) and c(iii). Parts of d were presented in \cref{fig:3}.   }\label{ED7:couplings_full}
\end{figure*}

\begin{figure*}[ht!]
    \centering
    \includegraphics[width=\textwidth]{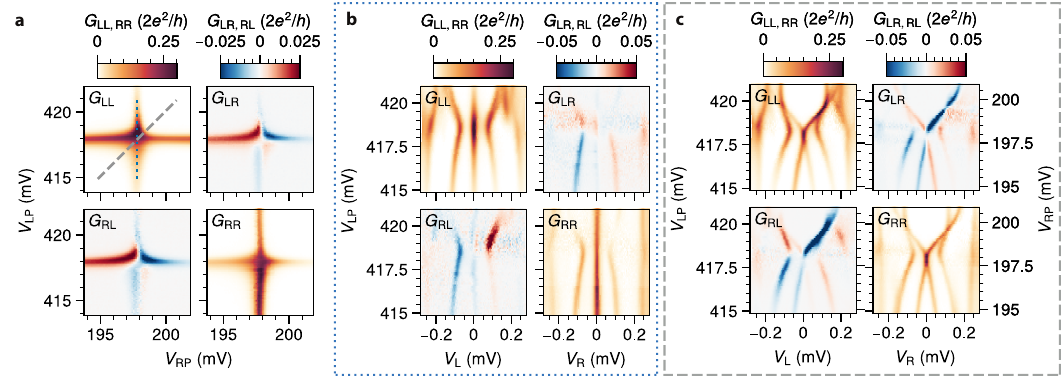}
    \caption{\textbf{Additional PMM spectra.} 
    \textbf{a,} CSD for a PMM sweet spot. Same as \cref{fig:4}a, but with different paths highlighted.
    \textbf{b,} PMM spectrum measured along the vertical path of panel a. The left QD is detuned, while the right one is kept on resonance. The asymmetry with respect to $\muL=0$ can be attributed to the asymmetric energy dispersion of the left YSR state (see \cref{ED3:QDs_sweetspot}a).
    \textbf{c,} PMM spectrum measured detuning both QDs along the diagonal path of panel a.
    The asymmetry with respect to $\muL=\muR=0$ can be explained as discussed in panel b.
    }\label{ED:PMM_extrapaths}
\end{figure*}

\begin{figure*}[ht!]
    \vspace{-1cm}
    \centering
    \includegraphics[width=\textwidth]{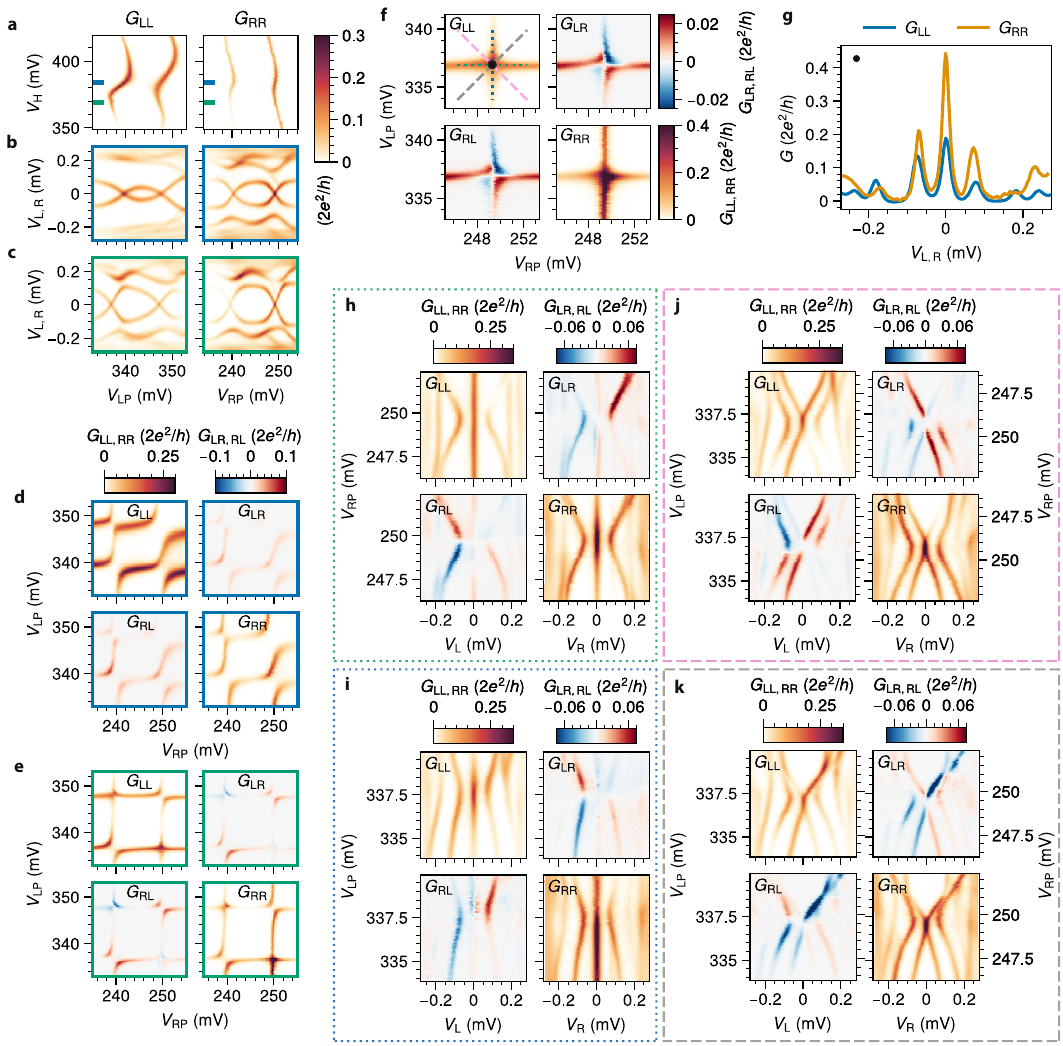}
    \caption{\textbf{PMM with different YSR and ABSs.}
    \textbf{a--c} Similarly to \cref{fig:1}, QD-ABS CSDs (a), spectra of the YSR states when the ABS is at its minimum energy (b), and for a $\VH$ corresponding to a $t = \Delta$ sweet spot (c), as described in the next panels. The colors of the frames correspond to the $\VH$ values indicated in panel a.
    \textbf{d--e,} Conductance matrix for YSR states CSDs when the ABS is at its minimum energy (d) and slightly detuned (e), as discussed in \cref{fig:3}. The color of the frames correspond to the $\VH$ values indicated in panel a.
    \textbf{f,} Zoom-in on the bottom right crossing of panel e, corresponding to a PMM sweet spot.
    \textbf{g,} Left and right conductance spectra measured at the PMM sweet spot, corresponding to the black dot at the middle of the crossing in \textbf{f}. The gap between the ground and the first excited states is $\Egap \approx \SI{71}{\micro eV}$. 
    \textbf{h--k,} Conductance matrix of the PMM spectra along the paths in panel f, analogously to \cref{fig:4} and \cref{ED:PMM_extrapaths}.
    }\label{ED:PMM2}
\end{figure*}

\begin{figure*}[ht!]
    \centering
    \includegraphics[width=\textwidth]{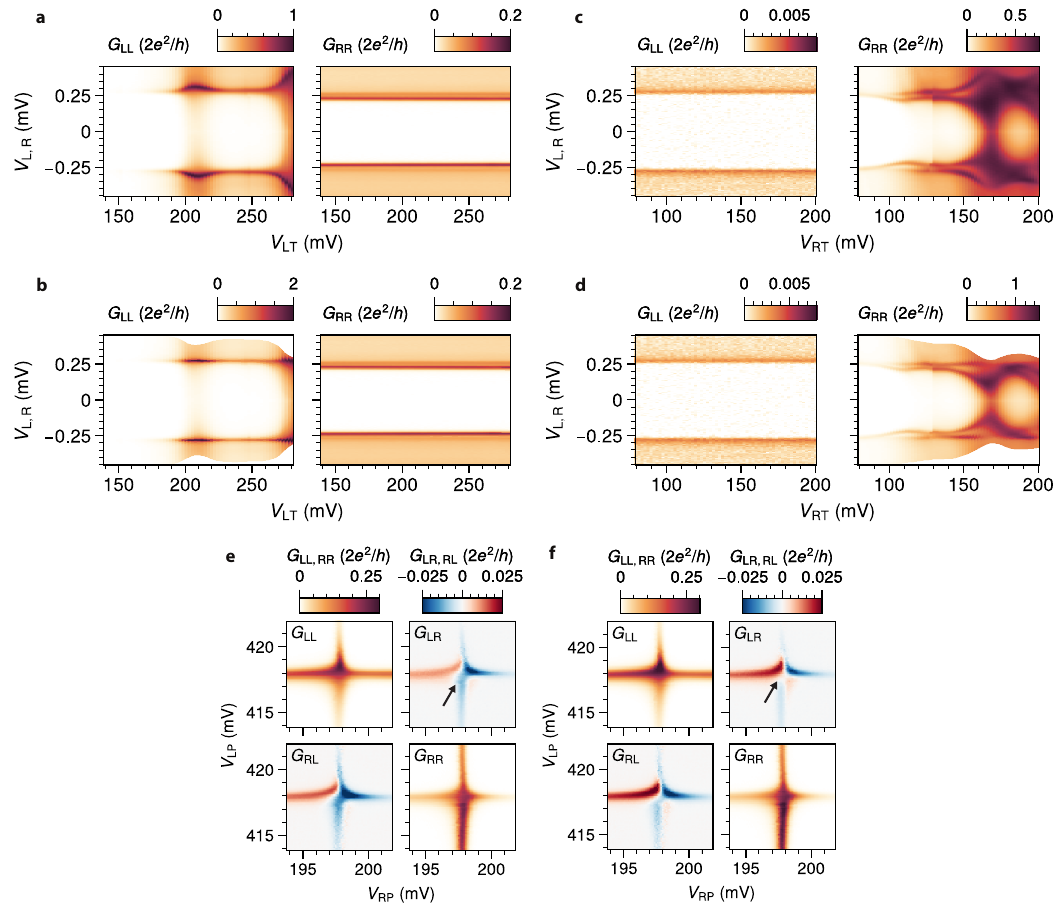}
    \caption{\textbf{Estimation of series resistance.} 
    \textbf{a,} Tunnel spectroscopy of the hybrid spectrum from both sides varying $\VLI$. Whenever there is finite subgap conductance, the parent gap expands because of the voltage divider effect.
    \textbf{b,} Same measurement as in panel a, after correcting the voltage divider effect. We find that a series resistance of $R_\mathrm{series} = \SI{3.65}{k\Omega}$ corrects the gap increase discussed above. In addition, we also considered a total resistance for the voltage source and current meter of $\SI{300}{\Omega}$.
    \textbf{c, d,} Same as panels a and b, but for the right side.
    \textbf{e, f,} CSD presented in \cref{fig:4}a, before (e) and after (f) correcting the voltage divider effect. The local signal is not significantly affected. On the other hand, the alternating positive and negative non-local signal becomes clearer after the correction. In particular, it approaches zero in the middle of the crossing, indicated by the black arrow.
    Without series resistance correction, the finite non-local signal at the center of the crossing in panel e could be misinterpreted as a signature of low Majorana polarization \cite{Tsintzis.2022}.
    }\label{ED8:Rseries}
\end{figure*}

\renewcommand\thefigure{ED\arabic{figure}}
\setcounter{figure}{0}

\end{document}